\documentclass[aip,reprint,nofootinbib]{revtex4-1}
\usepackage{graphicx}
\usepackage{bm}

\begin{document}
\title{Analytic Expressions for Exponentials of Specific Hamiltonian Matrices}
\date{\today}
\author{C. Baumgarten}
\affiliation{Paul Scherrer Institute, Switzerland}
\email{christian.baumgarten@psi.ch}

\def\begeq{\begin{equation}}
\def\endeq{\end{equation}}
\def\begary{\begeq\begin{array}}
\def\endary{\end{array}\endeq}
\def\bmtx{\left(\begin{array}}
\def\emtx{\end{array}\right)}
\def\eps{\varepsilon}
\def\d{\partial}
\def\y{\gamma}
\def\w{\omega}
\def\W{\Omega}
\def\s{\sigma}
\def\ket#1{\left|\,#1\,\right>}
\def\bra#1{\left<\,#1\,\right|}
\def\bracket#1#2{\left<\,#1\,\vert\,#2\,\right>}
\def\erw#1{\left<\,#1\,\right>}

\def\Exp#1{\exp\left(#1\right)}
\def\Log#1{\ln\left(#1\right)}
\def\Sinh#1{\sinh\left(#1\right)}
\def\Sin#1{\sin\left(#1\right)}
\def\Tanh#1{\tanh\left(#1\right)}
\def\Tan#1{\tan\left(#1\right)}
\def\Cos#1{\cos\left(#1\right)}
\def\Cosh#1{\cosh\left(#1\right)}

\begin{abstract}
Hamiltonian matrices appear in a variety or problems in physics and engineering, mostly
related to the time evolution of linear dynamical systems as for instance in ion beam optics.
The time evolution is given by symplectic transfer matrices which are the exponentials of the
corresponding Hamiltonian matrices. We describe a method to compute analytic formulas for 
the matrix exponentials of Hamiltonian matrices of dimensions $4\times 4$ and $6\times 6$. 
The method is based on the Cayley-Hamilton theorem and the Faddeev-LeVerrier method to 
compute the coefficients of the characteristic polynomial. The presented method 
is extended to the solutions of $2\,n\times 2\,n$-matrices when the roots of the 
characteristic polynomials are computed numerically. The main advantage of this method 
is a speedup for cases in which the exponential has to be computed for a number of different 
points in time or positions along the beamline.
\end{abstract}
%PACS-codes:
%Hamiltonian Mechanics,  45.20.Jj, 47.10.Df
%Coupled Oscillators, 05.45.Xt
\pacs{45.20.Jj, 47.10.Df, 05.45.Xt}
\keywords{Hamiltonian mechanics, Coupled Oscillators}
\maketitle

\section{Introduction}

Hamiltonian matrices are often derived from the general classical oscillator with $n$
coupled degrees of freedom, for instance in linear coupled (ion beam) optics. 
Let $\psi=(q_1,p_1,\dots,q_n,p_n)^T$ be the state vector 
of a classical dynamical system with $n$ degrees of freedom, where $q_i$ are the canonical 
coordinates and $p_i$ the canonical momenta with the Hamiltonian function ${\cal H}$ given by~\footnote{
An introduction into linear Hamiltonian theory can be found in Meyer, Offin and Hall~\cite{MHO}.}
\begeq
{\cal H}=\frac{1}{2}\,\psi^T\,{\cal A}\,\psi
\endeq
with the symmetric matrix ${\cal A}$, then the Hamiltonian equations of motion can be written as
\begeq
\dot\psi=\y_0\,\nabla_{\psi}\,{\cal H}=\y_0\,{\cal A}\,\psi={\bf F}\,\psi\,,
\endeq
where the overdot indicates the derivative with respect to a time-like variable,
$\nabla\psi$ is the phase space gradient and $\y_0$ is the so-called symplectic unit matrix:
\begeq
{\y_0}^{(n)}=\mathrm{Diag}(\eta,\dots,\eta)\,.
\endeq
with $n$ blocks of size $2\times2$
\begeq
\eta=\bmtx{cc}0&1\\-1&0\emtx\,.
\endeq
In the following we skip the dimensional indicator and simply write $\y_0$ for
the symplectic unit matrix of any dimension and $\eta$, if we explicitely refer 
to $\y_0^{(1)}$. The matrix ${\bf F}=\y_0\,{\cal A}$ is called Hamiltonian and holds
\begeq
{\bf F}=\y_0\,{\bf F}^T\,\y_0\,.
\endeq
The system 
\begeq
\dot\psi={\bf F}\,\psi\,,
\endeq
has the straightforward solution
\begeq
\psi(\tau)=\exp{({\bf F}\,\tau)}\,\psi(0)\,.
\endeq
The matrix exponential (the ``transfer matrix'')
\begeq
{\bf M}(\tau)=\exp{({\bf F}\,\tau)}
\endeq
is symplectic since it can be shown that
\begeq
{\bf M}^T\,\y_0\,{\bf M}=\y_0\,.
\endeq
The matrix exponential of ${\bf F}$ can be computed by various methods, a crtitical overview can be found 
in Ref.~(\cite{w19,w19x}). In Ref.~(\cite{geo_paper}) we described a straightforward method to determine a 
sequence of symplectic transformations ${\bf R}_k$ that transforms $4\times 4$ Hamiltonian matrices with 
real, imaginary or zero eigenvalues to normal form, which can be applied iteratively to $2\,n\times 2\,n$ 
Hamiltonian matrices. 

The normal form is given by
\begeq
{\bf F}=\mathrm{Diag}(\w_1\,\eta,\w_2\,\eta,\dots,\w_n\,\eta)\,.
\endeq
Then the matrix exponential can directly be solved by blockwise exponentiation using Euler's formula~\footnote{
Note that $\eta^2=-{\bf 1}$ and hence $\eta$ is a representation of the unit imaginary. 
}:
\begeq
\exp{(\w\,\eta\,\tau)}={\bf 1}\,\cos{(\w\,\tau)}+\eta\,\sin{(\w\,\tau)}\,.
\endeq
After the exponentiation has been done, one applies the inverse symplectic transformation to obtain
the solution in the original coordinates. The method to use symplectic transformations has the advantage
that it can be applied to all Hamiltonian matrices with zero, real or imaginary eigenvalues without
restriction, that it is numerically stable and that it does not only allow to compute the matrix 
exponential, but yields the eigenvalues (and eigenvectors, if required~\cite{geo_paper}) as well.

However, there are alternative methods that might be superior, if the complete information given by 
the symplectic decoupling transformation is not used or if the problem is large and the decoupling
numerically too expensive. A simple method specifically for small values of $\tau$ would be the direct 
evaluation of the (truncated) exponential series
\begeq
{\bf M}(\tau)=\sum\limits_{k=0}^{k_{max}}\,{{\bf F}^k\,\tau^k\over k!}
\label{eq_mtxexp}
\endeq
However, besides known stability issues~\cite{w19,w19x}, this method does (without further measures)
not ensure that the matrix ${\bf M}$ is {\it symplectic}, which results in changes of energy or emittance. 
Furthermore, the number of matrix multiplications for a given accuracy can be large and the accuracy 
depends on the value of $\tau$.

\section{Trace Operator and Eigenvalue Spectrum}

It is easy to verify that any odd power of a Hamiltonian matrix is
again Hamiltonian while every even power is skew-Hamiltonian:
\begary{rcl}
{\bf F}^{2k+1}&=&\y_0\,({\bf F}^{2k+1})^T\,\y_0\\
{\bf F}^{2k}&=&-\y_0\,({\bf F}^{2k})^T\,\y_0\\
\endary
Since Hamiltonian matrices are the product of a symmetric and a skew-symmetric matrix,
they have zero trace - and hence all odd powers have zero trace as well:
\begeq
\mathrm{Tr}({\bf F}^{2k+1})=0\,.
\endeq
As shown in Ref.~(\cite{MHO}), if $\lambda$ is an eigenvalue of a Hamiltonian matrix, 
then $-\lambda$ is also an eigenvalue. Hence the characteristic polynomial of a Hamiltonian
matrix has the form
\begeq
p(x)=\prod\limits_{j=1}^n\,(x-\lambda_j)\,(x+\lambda_j)=\prod\limits_{j=1}^n\,(x^2-\lambda_j^2)
\endeq
and the sums of powers of the eigenvalues can be obtained from:
\begeq
\mathrm{Tr}({\bf F}^{2k})=2\,\sum\limits_{j=1}^n\,\lambda_j^{2k}\,.
\endeq
According to the Cayley-Hamilton theorem any matrix solves its own characteristic equation. 
For an arbitary $2\,n\times 2\,n$-matrix ${\bf F}$ this implies that
\begeq
\sum_{k=0}^{2\,n}\,c_k\,{\bf F}^k=0\,.
\endeq
Hence the $2\,n$-th power of the matrix can be expressed as a linear combination of lower 
powers. Thus the matrix exponential for a matrix of size $2\,n\times 2\,n$ can
be written as:
\begeq
{\bf M}(\tau)=\sum\limits_{k=0}^{2n-1}\,x_k(\tau)\,{\bf F}^k\,.
\endeq
The problem is therefore solved by the determination of the coefficient functions $x_k(\tau)$.
As the time derivative of ${\bf M}$ is ${\bf M}\,{\bf F}$, one may write:
\begeq
{\bf\dot M}(\tau)=\sum\limits_{k=0}^{2n-1}\,x_k(\tau)\,{\bf F}^{k+1}\,,
\label{eq_dMdt1}
\endeq
and also
\begeq
{\bf\dot M}(\tau)=\sum\limits_{k=0}^{2n-1}\,{\dot x}_k(\tau)\,{\bf F}^k\,.
\label{eq_dMdt2}
\endeq
The highest matrix power in Eq.~\ref{eq_dMdt1} is then be replaced by the use of the Cayley-Hamilton
theorem and one obtains effectively a set of linear differential equations for the coefficient functions
$x_k(\tau)$.

\section{The Faddeev-LeVerrier Algorithm}

Let us express the eigenvalues $\lambda_k$ by $\lambda_k=i\,\w_k$ so that the
characteristic polynomial can be written as
\begeq
p(x)=\prod\limits_{k=1}^n\,(x^2+\w_k^2)
\endeq
The traces of the even matrix potentials allow to define $t_k$ according to
\begeq
t_k=(-1)^k\,\frac{1}{2}\,\mathrm{Tr}({\bf F}^{2k})=\sum\limits_{j=1}^n\,\w_j^{2k}
\label{eq_traces}
\endeq
such that
\begary{rcl}
t_1&=&\sum\limits_{j=1}^n\,\w_j^2\\
t_2&=&\sum\limits_{j=1}^n\,\w_j^4\\
t_3&=&\sum\limits_{j=1}^n\,\w_j^6\\
&\vdots&\,.
\endary
Now we define the following sequence:
\begary{rcl}
p_0&=&1\\
p_1&=&t_1\\
p_{n+1}&=&\frac{1}{n+1}\,\sum\limits_{k=0}^n\,(-1)^k\,p_{n-k}\,t_{k+1}\\
\label{eq_FLV}
\endary
such that
\begary{rcl}
p_2&=&(p_1\,t_1-p_0\,t_2)/2\\
p_3&=&(p_2\,t_1-p_1\,t_2+p_0\,t_3)/3\\
p_4&=&(p_3\,t_1-p_2\,t_2+p_1\,t_3-p_0\,t_4)/4\\
&\vdots&\\
\endary
Then the characteristic polynomial $p(x)$ of the matrix ${\bf F}$ is
\begeq
p(x)=\sum_{k=0}^n\,x^{2k}\,p_{n-k}\,
\label{eq_cpoly}
\endeq 
This is known as Faddeev-LeVerrier algorithm~\cite{FLWiki,Hou,Helmberg}.

In the case of $4\times 4$ Hamiltonian matrices with two pairs of eigenvalues
one obtains for instance:
\begary{rcl}
t_1&=&\w_1^2+\w_2^2\\
t_2&=&\w_1^4+\w_2^4\\
p_0&=&1\\ 
p_1&=&t_1\\
p_2&=&(p_1\,t_1-p_0\,t_2)/2\\
   &=&((\w_1^2+\w_2^2)^2-(\w_1^4+\w_2^4))/2\\
   &=&\w_1^2\,\w_2^2\\
\label{eq_4x4}
\endary
The polynomial then is
\begary{rcl}
0&=&x^4+p_1\,x^2+p_2\\
0&=&x^4+(\w_1^2+\w_2^2)\,x^2+\w_1^2\,\w_2^2\\
0&=&(x^2+\w_1^2)\,(x^2+\w_2^2)\\
\endary
such that the eigenvalues are $\pm\,i\,\w_1$ and $\pm\,i\,\w_1$, as expected.

In case of dimension $6\times 6$, we find
\begary{rcl}
t_1&=&\w_1^2+\w_2^2+\w_3^2\\
t_2&=&\w_1^4+\w_2^4+\w_3^4\\
t_3&=&\w_1^6+\w_2^6+\w_3^6\\
p_0&=&1\\ 
p_1&=&t_1=\w_1^2+\w_2^2+\w_3^2\\
p_2&=&(p_1\,t_1-p_0\,t_2)/2\\
   &=&\w_1^2\,\w_2^2+\w_2^2\,\w_3^2+\w_1^2\,\w_3^2\\
p_3&=&(p_2\,t_1-p_1\,t_2+p_0\,t_3)/3\\
   &=&\w_1^2\,\w_2^2\,\w_3^2\\
\label{eq_6x6}
\endary
If we insert these coefficients into Eq.~\ref{eq_cpoly},
it is easily seen that we again find the characteristic polynomial.
Obviously the coefficient $p_1$ equals the sum
\begeq
p_1=\sum\limits_{k=1}^n\,\w_k^{2}\,,
\endeq
and $p_n$ the product of all squared eigenfrequencies:
\begeq
p_n=\prod\limits_{k=1}^n\,\w_k^{2}\,.
\endeq
Hence the matrix ${\bf F}$ is regular, if $p_n\ne 0$. If ${\bf F}$ has two
vanishing pairs of eigenvalues, then the last two coefficients vanish, 
$p_n=p_{n-1}=0$, and so on.

Hence the Faddeev-LeVerrier evaluation of the traces of the matrix monomials 
allows not only to obtain the characteristic polynomial, but also to determine
the number of non-zero eigenvalues of the matrix ${\bf F}$, i.e. to decide
whether the matrix is singular.

\section{$4\times 4$-Matrices}
\label{sec_4x4}

In the following we show how the method can be applied to the (important)
special case of $4\times 4$ Hamiltonian matrices. There are two pairs of 
eigenvalues $\pm i\,\w_1$ and $\pm i\,\w_2$. If these eigenvalues are distinct,
then the characteristic polynomial is (Eqs.~\ref{eq_cpoly},\ref{eq_4x4}):
\begeq
x^4+p_1\,x^2+p_2=0
\endeq
with
\begary{rcl}
p_1&=&\w_1^2+\w_2^2\\
p_2&=&\w_1^2\,\w_2^2\,.
\label{eq_a_b_def}
\endary
Multiplication of the first Eq.~\ref{eq_a_b_def} with either $\w_1^2$ or $\w_2^2$ gives~\footnote{
As we had chosen before to write the eigenvalues as $\lambda_k=i\,\w_k$, we obtain a sign change
of the k-th power with $k\,\mathrm{ mod }\,4 = 2$.
}:
\begary{rcl}
\w_1^2\,p_1&=&\w_1^4+\w_2^2\,\w_1^2\\
\w_2^2\,p_1&=&\w_2^2\,\w_1^2\,+\w_2^4\\
0&=&\w^4-\w^2\,p_1+p_2\\
\endary
The frequencies are then given by
\begary{rcl}
\w_1&=&\pm\sqrt{\frac{p_1}{2}+\sqrt{\frac{p_1^2}{4}-p_2}}\\
\w_2&=&\pm\sqrt{\frac{p_1}{2}-\sqrt{\frac{p_1^2}{4}-p_2}}\\
\label{eq_freq}
\endary
and the characteristic equation of ${\bf F}$ yields
\begeq
{\bf F}^4=-{\bf F}^2\,p_1-p_2\,.
\endeq
The time derivatives of ${\bf M}$ give, according to Eqs.~(\ref{eq_dMdt1}) 
and (\ref{eq_dMdt2}):
\begary{rcl}
{\bf\dot M}&=&{\bf M}\,{\bf F}\\
           &=&x_0\,{\bf F}+x_1\,{\bf F}^2+x_2\,{\bf F}^3+x_3\,{\bf F}^4\\
           &=&x_0\,{\bf F}+x_1\,{\bf F}^2+x_2\,{\bf F}^3-x_3\,(p_2+p_1\,{\bf F}^2)\\
           &=&-p_2\,x_3+x_0\,{\bf F}+(x_1-p_1\,x_3)\,{\bf F}^2+x_2\,{\bf F}^3\\
{\bf\dot M}&=&\dot x_0+{\dot x}_1\,{\bf F}+{\dot x}_2\,{\bf F}^2+\dot x_3\,{\bf F}^3\\
\endary
so that
\begary{rcl}
{\dot x}_0&=&-p_2\,x_3\\
{\dot x}_1&=&x_0\\
{\dot x}_2&=&x_1-p_1\,x_3\\
{\dot x}_3&=&x_2\\
\label{eq_exp_coeffs}
\endary
or, written with ${\bf x}=(x_0,x_1,x_2,x_3)^T$ in matrix form:
\begary{rcl}
{\bf\dot x}&=&{\bf G}\,{\bf x}\\
\bmtx{c}
{\dot x}_0\\
{\dot x}_1\\
{\dot x}_2\\
{\dot x}_3\\
\emtx&=&\bmtx{cccc}
0&0&0&-p_2\\
1&0&0&0\\
0&1&0&-p_1\\
0&0&1&0\\
\emtx\,\bmtx{c}
x_0\\
x_1\\
x_2\\
x_3\\
\emtx
\label{eq_diff4x4}
\endary
This equation could again be solved by the matrix exponential - of ${\bf G}$ - so it 
does not seem that we gained much. However, the number of variables is now reduced from $10$ 
in ${\bf F}$ to $2$, namely to $p_1$ and $p_2$ in ${\bf G}$, and the matrix form of ${\bf G}$
allows for a (more or less) direct solution. 
Note that the matrix ${\bf G}$ fulfills the same characteristic equation as ${\bf F}$ and 
therefore has the same eigenvalues, i.e. is similar to ${\bf F}$. Furthermore we know
from Eq.~\ref{eq_mtxexp} in combination with the characteristic equation, that 
$x_0$ and $x_2$ are even functions of $\tau$ while $x_1$ and $x_3$ are odd, so that
\begary{rcl}
x_0(-\tau)&=&x_0(\tau)\\
x_1(-\tau)&=&-x_1(\tau)\\
x_2(-\tau)&=&x_2(\tau)\\
x_3(-\tau)&=&-x_3(\tau)\\
\endary
and from ${\lim\atop\tau\to 0}{\bf M}(\tau)={\bf 1}$ we have ${\bf x}(0)=(1,0,0,0)^T$.
Therefore we make the following Ansatz such that the second and fourth of Eq.~\ref{eq_exp_coeffs}
are already fulfilled:
\begary{rcl}
x_0(\tau)&=&x_0^{(1)}\,\cos{(\w_1\,\tau)}+x_0^{(2)}\,\cos{(\w_2\,\tau)}\\
x_1(\tau)&=&{x_0^{(1)}\over\w_1}\,\sin{(\w_1\,\tau)}+{x_0^{(2)}\over\w_2}\,\sin{(\w_2\,\tau)}\\
x_2(\tau)&=&x_2^{(1)}\,\cos{(\w_1\,\tau)}+x_2^{(2)}\,\cos{(\w_2\,\tau)}\\
x_3(\tau)&=&{x_2^{(1)}\over\w_1}\,\sin{(\w_1\,\tau)}+{x_2^{(2)}\over\w_2}\,\sin{(\w_2\,\tau)}\\
\endary
The remaining equations are fulfilled, if $x_0^{(1)}=\w_2^2\,x_2^{(1)}$ and 
$x_0^{(2)}=\w_1^2\,x_2^{(2)}$. The starting condition ${\bf x}(0)=(1,0,0,0)^T$ requires
that $x_0^{(1)}+x_0^{(2)}=1$ and $x_2^{(1)}=-x_2^{(2)}=x_2$ so that we finally obtain:
\begary{rcl}
x_0^{(2)}&=&\w_1^2\,x_2\\
x_0^{(1)}&=&-\w_2^2\,x_2\\
(\w_1^2-\w_2^2)\,x_2&=&1\\
x_0^{(2)}&=&{\w_1^2\over\w_1^2-\w_2^2}\\
x_0^{(1)}&=&-{\w_2^2\over\w_1^2-\w_2^2}\\
x_2^{(1)}&=&-{1\over\w_1^2-\w_2^2}\\
x_2^{(2)}&=&{1\over\w_1^2-\w_2^2}\\
\endary
The conditions $x_1(0)=0$ and $x_3(0)=0$ are automatically fulfilled.
Hence we can compute the matrix exponential of ${\bf F}$ by computing the trace of 
${\bf F}^2$ and ${\bf F}^4$ and solving a second order polynomial.

The solution can be generalized straightforward to include real eigenvalues (e.g. imaginary
frequencies) by the use of the relations
\begary{rcl}
\sin{(i\,x)}&=&i\,\sinh{(x)}\\
\cos{(i\,x)}&=&\cosh{(x)}\\
\endary

\subsection{Degenerate $4\times 4$-Matrices}

If the system is degenerate $\w_1=\w_2=\w\ne 0$, then one might think
that we obtain Eq.~\ref{eq_diff4x4} and Eq.~\ref{eq_a_b_def}:
\begary{rcl}
{\bf\dot x}&=&\bmtx{cccc}
0&0&0&-\w^4\\
1&0&0&0\\
0&1&0&-2\,\w^2\\
0&0&1&0\\
\emtx\,{\bf x}
\endary
This, however, is wrong: The square of the degenerate $4\times 4$-Matrix ${\bf F}$ is
proportional to the unit matrix
\begary{rcl}
{\bf F}&=&{\bf E}\,\mathrm{Diag}(i\,\w,-i\,\w,i\,\w,-i\,\w){\bf E}^{-1}\\
{\bf F}^2&=&{\bf E}\,\mathrm{Diag}(-\w^2,-\w^2,-\w^2,-\w^2){\bf E}^{-1}=-\w^2\,{\bf 1}\\
\endary
such that the $4\times 4$ problem ``collapses'' and reduces
effectively to the case of a $2\times 2$-matrix so that
\begary{rcl}
{\bf M}(\tau)={\bf 1}\,\cos{(\w\,\tau)}+{\bf F}/\w\,\sin{(\w\,\tau)}\,.
\endary

\subsection{Singular $4\times 4$-Matrices}

If both eigenvalues vanish, then $p_1=p_2=0$ and the solution of Eq.~\ref{eq_diff4x4}
is readily solved by direct integration. In combination with the boundary and symmetry 
conditions it follows that
\begary{rcl}
x_0&=&1\\
x_1&=&\tau\\
x_2&=&\tau^2/2\\
x_3&=&\tau^3/6\,,
\endary
such that the ``truncated power series'' is the exact solution: 
\begeq
{\bf M}(\tau)={\bf 1}+{\bf F}\,\tau+{\bf F}^2\,\tau^2/2+{\bf F}^3\,\tau^3/6\,.
\endeq
This special case in which all eigenvalues vanish, can immediately be generalized to
any matrix dimension.

If one of the two eigenvalues is zero, then $p_2=0$ but $p_1=\w^2\ne 0$ and the solution of 
Eq.~\ref{eq_diff4x4} is a mixture of both cases:
\begary{rcl}
\bmtx{c}
{\dot x}_0\\
{\dot x}_1\\
{\dot x}_2\\
{\dot x}_3\\
\emtx&=&\bmtx{cccc}
0&0&0&0\\
1&0&0&0\\
0&1&0&-\w^2\\
0&0&1&0\\
\emtx\,\bmtx{c}
x_0\\
x_1\\
x_2\\
x_3\\
\emtx
\label{eq_mix4x4}
\endary
so that the integration results in accordance with the boundary 
conditions ${\bf x}(0)=(1,0,0,0)^T$:
\begary{rcl}
x_0&=&1\\
x_1&=&\tau\\
x_2&=&c_0\,(\cos{(\w\,\tau)}-1)\\
x_3&=&{\tau\over\w^2}+c_1\,\sin{(\w\,\tau)}\,.
\endary
The constants are obtained from the third row of Eq.~\ref{eq_mix4x4}:
\begary{rcl}
{\dot x}_2&=&-c_0\,\w\,\sin{(\w\,\tau)}\\
&=&x_1-\w^2\,x_3\\
&=&-c_1\,\w^2\,\sin{(\w\,\tau)})
\endary
so that
\begary{rcl}
c_0&=&\w\,c_1\\
\endary
And finally from ${\dot x}_3=x_2$ it follows that $c_1=-\frac{1}{\w^3}$.
Hence the solution is
\begeq
{\bf M}(\tau)={\bf 1}+{\bf F}\,\tau+{1-\cos{(\w\,\tau)}\over\w^2}\,{\bf F}^2+{\tau\w-\sin{(\w\,\tau)}\over\w^3}\,{\bf F}^3\,.
\endeq

\section{Matrix Exponential for Sp(6)}

Consider a non-singular and non-degenerate Hamiltonian matrix ${\bf F}$ of dimension $6\times 6$; 
the coefficients of the characteristic equation are given in Eq.~\ref{eq_6x6}.
The Caley-Hamilton theorem can be expressed as
\begeq
{\bf F}^6=-p_1\,{\bf F}^4-p_2\,{\bf F}^2-p_3\,.
\endeq
The matrix exponential can therefore be expressed by six terms:
\begary{rcl}
{\bf M}&=&\sum\limits_{k=0}^5\,x_k(\tau)\,{\bf F}^k\\
{\bf \dot M}&=&\sum\limits_{k=0}^5\,{\dot x}_k(\tau)\,{\bf F}^k\\
{\bf \dot M}&=&{\bf M}\,{\bf F}=\sum\limits_{k=0}^5\,x_k(\tau)\,{\bf F}^{k+1}\\
\endary
so that with
\begary{rcl}
{\dot x}_0&=&-x_5\,p_3\\
{\dot x}_1&=&x_0\\
{\dot x}_2&=&x_1-x_5\,p_2\\
{\dot x}_3&=&x_2\\
{\dot x}_4&=&x_3-x_5\,p_1\\
{\dot x}_5&=&x_4\\
\endary
one obtains the equation system:
\begary{rcl}
{\bf \dot x}&=&{\bf G}\,{\bf x}\\
{\bf G}&=&\bmtx{cccccc}
0&0&0&0&0&-p_3\\
1&0&0&0&0&0\\
0&1&0&0&0&-p_2\\
0&0&1&0&0&0\\
0&0&0&1&0&-p_1\\
0&0&0&0&1&0\\
\emtx
\label{eq_sp6_g}
\endary
The construction of the system is such that for a known $x_4$ the
remaining coefficients can be computed straightforward {\it if} the system 
is purely oscillatory as assumed in the previous section.
\begary{rcl}
x_4(\tau)&=&\sum\limits_k\,x_4^{(k)}\,\cos{(\w_k\,\tau)}\\
x_5(\tau)&=&\sum\limits_k\,x_4^{(k)}\,{\sin{(\w_k\,\tau)}\over\w_k}\\
x_0(\tau)&=&\sum\limits_k\,x_4^{(k)}\,{p_3\over\w_k^2}\,\cos{(\w_k\,\tau)}\\
x_1(\tau)&=&\sum\limits_k\,x_4^{(k)}\,{p_3\over\w_k^3}\,\sin{(\w_k\,\tau)}\\
x_2(\tau)&=&\sum\limits_k\,x_4^{(k)}\,(-{p_3-p_2\,\w_k^2\over\w_k^4})\,\cos{(\w_k\,\tau)}\\
x_3(\tau)&=&\sum\limits_k\,x_4^{(k)}\,(-{p_3-p_2\,\w_k^2\over\w_k^5})\,\sin{(\w_k\,\tau)}\\
x_4(\tau)&=&\sum\limits_k\,x_4^{(k)}\,({p_3-p_2\,\w_k^2+p_1\,\w_k^4\over\w_k^6})\,\cos{(\w_k\,\tau)}\\
\label{eq_even_coeffs}
\endary
The boundary conditions are now:
\begary{rcl}
x_0(0)&=&\sum\limits_k\,x_4^{(k)}\,{p_3\over\w_k^2}=1\\
x_2(0)&=&\sum\limits_k\,x_4^{(k)}\,(-{p_3-p_2\,\w_k^2\over\w_k^4})=0\\
x_4(0)&=&\sum\limits_k\,x_4^{(k)}\,({p_3-p_2\,\w_k^2+p_1\,\w_k^4\over\w_k^6})=0\,,
\label{eq_bondary_sp6}
\endary
which can be written in matrix form as
\begeq
{\bf P}\,\bmtx{c}
x_4^{(1)}\\
x_4^{(2)}\\
x_4^{(3)}\\
\emtx=\bmtx{c}1\\0\\0\emtx
\endeq
where
\begeq
{\bf P}=
\bmtx{ccc}
{p_3\over\w_1^2}&{p_3\over\w_2^2}&{p_3\over\w_3^2}\\
{p_3-p_2\,\w_1^2\over\w_1^4}&{p_3-p_2\,\w_2^2\over\w_2^4}&{p_3-p_2\,\w_3^2\over\w_3^4}\\
{p_3-p_2\,\w_1^2+p_1\,\w_1^4\over\w_1^6}&{p_3-p_2\,\w_2^2+p_1\,\w_2^4\over\w_2^6}&{p_3-p_2\,\w_3^2+p_1\,\w_3^4\over\w_3^6}\\
\emtx\,.
\endeq
If one replaces $p_k$ with the expressions of Eq.~\ref{eq_6x6}, one obtains
\begary{rcl}
x_4^{(1)}&=&{1\over(\w_1^2-\w_2^2)(\w_1^2-\w_3^2)}\\
x_4^{(2)}&=&{1\over(\w_2^2-\w_1^2)(\w_2^2-\w_3^2)}\\
x_4^{(3)}&=&{1\over(\w_3^2-\w_1^2)(\w_3^2-\w_2^2)}\\
\endary

\subsection{Singular $6\times 6$ Matrices}

In case of a single vanishing Eigenvalue $\w_3=0$ of $6\times 6$ Matrices, one obtains the following
coefficients:
\begary{rcl}
x_0(\tau)&=&1\\
x_1(\tau)&=&\tau\\
x_2(\tau)&=&{(\cos{(\w_1\,\tau)}-1)\,\w_2^4-(\cos{(\w_2\,\tau)}-1)\,\w_1^4\over\w_1^2\,\w_2^2\,(w_1^2-w_2^2)}\\
x_3(\tau)&=&{(\sin{(\w_1\,\tau)}-\w_1\,\tau)\,\w_2^5-(\sin{(\w_2\,\tau)}-\w_2\,\tau)\,\w_1^5\over\w_1^3\,\w_2^3\,(w_1^2-w_2^2)}\\
x_4(\tau)&=&{(\cos{(\w_1\,\tau)}-1)\,\w_2^2-(\cos{(\w_2\,\tau)}-1)\,\w_1^2\over\w_1^2\,\w_2^2\,(w_1^2-w_2^2)}\\
x_5(\tau)&=&{(\sin{(\w_1\,\tau)}-\w_1\,\tau)\,\w_2^3-(\sin{(\w_2\,\tau)}-\w_2\,\tau)\,\w_1^3\over\w_1^3\,\w_2^3\,(w_1^2-w_2^2)}\\
\label{eq_6x6a}
\endary

In case of two vanishing Eigenvalues $\w_2=\w_3=0$ and $\w_1=\w$, the coefficients are given by
\begary{rcl}
x_0(\tau)&=&1\\
x_1(\tau)&=&\tau\\
x_2(\tau)&=&\tau^2/2\\
x_3(\tau)&=&\tau^3/6\\
x_4(\tau)&=&{\cos{(\w\,\tau)}-1\over\w^4}+{\tau^2\over 2\,\w^2}\\
x_5(\tau)&=&{\sin{(\w\,\tau)}-\w\,\tau\over \w^5}+{\tau^3\over 6\,\w^2}\\
\label{eq_6x6a}
\endary

\section{Matrix Exponential for Sp(2n)}

The generalization of $Sp(6)$ to $Sp(2n)$ is straightforward for the non-singular 
(non-degenerate) case and can be summarized as follows:
\begin{enumerate}
\item Compute the $2\,n$ matrix powers ${\bf F}^k$ with $k\in\,[1\dots 2\,n]$.
\item Compute the $n$ matrix traces to determinne $t_k$ according to Eq.~\ref{eq_traces}.
\item Use Faddeev-LeVerrier method according to Eq.~\ref{eq_FLV} to obtain $n$ coefficients $p_k$ 
of the characteristic polynomial.
\item Compute the $n$ eigenvalues as roots of the characteristic polynomial by known
numerical methods~\cite{Bairstow}.
\item $x_{2n-2}$ can be computed from the eigenvalues $\lambda_k=i\,\w_k$ according to
$$x_{(2n-2)}(\tau)=\sum\limits_{k=1}^n\,\left(\prod\limits_{j\ne k}\,{1\over\w_k^2-\w_j^2}\right)\,\cos{(\w_k\,\tau)}$$
\item Solve the remaining terms of Eq.~\ref{eq_spn_g} as described below.
\end{enumerate}
The remaining coefficient functions ($x_0(\tau)\dots x_{2n-1}(\tau)$) are:
\begary{rcl}
x_{(2n-1)}(\tau)&=&\sum\limits_{k=1}^n\,\left(\prod\limits_{j\ne k}\,{1\over\w_k(\w_k^2-\w_j^2)}\right)\,\sin{(\w_k\,\tau)}\\
x_{(2n-2k)}&=&\dot x_{(2n-2k+1)}\\
x_{(2n-2k-1)}&=&\dot x_{(2n-2k)}+p_k\,x_{(2n-1)}\\
\endary
which solves the system
\begary{rcl}
{\bf \dot x}&=&{\bf G}\,{\bf x}\\
{\bf G}&=&\bmtx{ccccccc}
0&0&0&      &0&0&-p_n\\
1&0&0&\dots &0&0&0\\
0&1&0&      &0&0&-p_{n-1}\\
0&0&1&      &0&0&0\\
 &&\vdots&&   &\vdots&\\
0&0&0&\dots&1&0&-p_1\\
0&0&0&     &0&1&0\\
\emtx
\label{eq_spn_g}
\endary

\begin{enumerate}
\item Compute the $2\,n$ matrix powers ${\bf F}^k$ with $k\in\,[1\dots 2\,n]$.
\item Compute the $n$ matrix traces to determinne $t_k$ according to Eq.~\ref{eq_traces}.
\item Use Faddeev-LeVerrier method according to Eq.~\ref{eq_FLV} to obtain $n$ coefficients $p_k$ 
of the characteristic polynomial.
\item Count the number of zero eigenvalue-pairs $m$: the first $2\,m$ functions $x_k(\tau)$ are the
monomials $x_k(\tau)={\tau^k\over k!}$ for $k\in\,[0\dots 2m-1]$.
\item If $n>m$, compute the $N=n-m$ non-zero eigenvalue pairs, e.g. the roots of the characteristic polynomial.
For $N\le 4$, this can be done directly, for $N>4$ this has to be done numerically.
\item Solve the remaining terms of Eq.~\ref{eq_spn_g} as described below,
\end{enumerate}

We gave some examples of how to solve Eq.~\ref{eq_spn_g}, but we did not yet give
a generally applicable solution for arbitary $n$. Consider the first steps have been
done, i.e. all $t_k$ and $p_k$ for $k\in\,[1\dots n]$ are known. Let $m\ge 0$ be the
number of vanishing $p_k$, i.e. $p_{n+1-j}=0$ for $j\in [1\dots m]$, then there are
$m$ eigenvalue pairs equal to zero.
Consider the {\it root}-solution for $x_{2n-1}(\tau)$ is written as:
\begeq
x_{2\,n-1}(\tau)=P_{2n-1}(\tau)+\sum\limits_k\,s_{2\,n-1}^{(k)}\,\sin{(\w_k\,\tau)}\,,
\endeq
where $s_j$ are the trigonometric coefficients and the polynome $P_{2n-1}(\tau)$ is 
odd and of order $2m-1$:
\begary{rcl}
P_{2n-1}(\tau)&=&\sum\limits_{k=1}^{m}\,c_{2n-1}^{(2k-1)}\,{\tau^{2\,k-1}\over (2\,k-1)!}\\
&=&c_{2n-1}^{(1)}\,\tau+c_{2n-1}^{(3)}\,\tau^3/6+\dots\\
\endary
Then, according to the the last row in Eq.~\ref{eq_spn_g}, we have
\begeq
x_{n-2}(\tau)={\dot x}_{n-1}\,,
\endeq
and subsequentially:
\begary{rcl}
x_{n-3}(\tau)&=&{\dot x}_{n-2}+p_1\,x_{n-1}\\
x_{n-4}(\tau)&=&{\dot x}_{n-3}\\
x_{n-5}(\tau)&=&{\dot x}_{n-4}+p_2\,x_{n-1}\\
&\vdots&\\
\endary
which can be summarized as follows:
\begary{rcl}
x_{n-2k}(\tau)&=&{\dot x}_{n-2k+1}\\
x_{n-2k-1}(\tau)&=&{\dot x}_{n-2k}+p_k\,x_{n-1}\\
\endary
so that
\begary{rcl}
x_{n-2k-1}(\tau)&=&{\ddot x}_{n-2k+1}+p_k\,x_{n-1}\\
\endary

\section{Conclusion}

We described a method that allows to compute the exponentials of Hamiltonian matrices.
For a matrix of size $2\,n\times 2\,n$, the method requires to compute the matrix powers
up to $2\,n-1$, to compute the traces of all even matrix powers, to generate the characteristic
polynomials with the Faddeev-LeVerrier-Algorithm using the traces, to compute the eigenvalues
of the characteristic polynomial and finally to solve for the coefficients $x_k(t)$.
The advantage of this method mainly is the speedup in the computation of the matrix exponential 
for various times $t_k$ or various positions along the beamline, respectively. 

%\begin{appendix}

%\end{appendix}

\end{document}